# Viewpoint: Journals for Certification, Conferences for Rapid Dissemination -- Rethinking the Role of Journals in Computer Science

Joseph Y. Halpern , David C. Parkes
August 3, 2010

The publication culture in Computer Science is different from that of all other disciplines. Whereas other disciplines focus on journal publication, the standard practice in CS has been to publish in a conference and then (sometimes) publish a journal version of the conference paper. Indeed, it is through publication in selective, leading conferences that the quality of CS research is typically assessed.

Why should a researcher publish a journal version of a paper? In fields other than CS, which place no value on conference publication, there are two main reasons:

> 1. Certification: Publication in a peer-reviewed journal is a signal to the world that the paper has passed a minimal level of competence; publication in a leading journal confers even more prestige.

> 2. Publicity: Journal publication can be an effective way to tell the world (including policy makers and science advisors, not just colleagues) about the research, particularly publication in a leading journal like Science or Nature.

In CS, the situation is different. In many subdisciplines of CS, having a paper accepted at a leading conference already gives as much of a certification as getting it into a leading journal. Conferences play the role of publicity as well. The best way to get your subdiscipline to know about your results is to publish them in the leading conference for that subdiscipline.

But there is increasing debate about the role of conferences in our field [1, 5, 8]. Fortnow [3] argues that our field suffers from the current use of conferences for certification.[1] Two particular problems that he cites are those of quality and innovation suffering because we end up living in a deadline-driven world," and the splintering of the field into multiple conferences (so that there are enough publication venues), leading to conferences failing to act as a broad forum and bring their communities together." Fortnow suggests that conferences should be held less frequently, and accept every reasonable paper for presentation without proceedings.

Even in the current situation, journals do play a role in providing more relaxed page limits, which allow authors to include more discussion, more expository details, details of proofs, additional experimental results, and the time to submit a more polished, thoughtful paper. For theoretical papers, the certification issue remains significant because it is rare that conference reviewers review proofs as thoroughly as journal reviewers. Publication in a journal also adds value through a paper going through a strict review process with several iterations. Finally, journals also provide publicity for interdisciplinary work.

We can already see the beginning of a shift in the conference and journal landscape. Part of the shift involves journals publishing conference proceedings as special issues. For example, the *ACM Transactions on Graphics* (*TOG*) publishes every SIGGRAPH and SIGGRAPH Asia technical paper in its biannual conference issues, which replace traditional conference proceedings. If a paper is conditionally accepted for presentation at SIGGRAPH, then the paper undergoes a second review (by one of the original reviewers) to ensure that all changes requested by the reviewers are made; it is then also published in *TOG*. Similarly, papers accepted at this year's International Conference on Logic Programming (ICLP) will appear in *Theory and Practice of Logic Programming*. However, note that this approach means that these journal issues suffer from the same problems that conference publications suffer from: papers are subject to the conference length restrictions and paper submission deadlines that make it difficult to do serious revisions.

The Database community has taken this one step farther. Currently the only way to submit a paper to the VLDB (Very Large Database) conference is to submit it to the journal *Proceedings of the VLDB Endowment* (*PVLDB*). Continuous submissions are accepted throughout the year, reviews guaranteed within 2 months, a full review cycle including checking of final versions by responsible editors is supported, and papers accepted by a specified date are offered a presentation slot in the next VLDB conference. This change makes is possible to have a "revise and resubmit review", but there still remain serious page limitations. (Interestingly, *PVLDB* also allows the publication of an extended version of a *PVLDB* paper in another journal.)

The *TOG*/SIGGRAPH relationship has another facet, which illustrates another possibility for a conference-journal relationship: *TOG* allows any author of a published paper to present the work at SIGGRAPH, while operating without deadlines and less strict page limits.

These experiments suggest that the CS community needs to think through the intertwined role of conferences and journals, especially in light of the growing amount of research at the intersection of computer science and other fields. Our unique distinction of being a conference-led field leads to a particular problem for multidisciplinary work because, outside of CS, journals typically have all the power, and are very reluctant to take papers where versions have appeared previously. For example, some leading Biology journals are unwilling to publish work that has appeared in RECOMB (Conference on Research in Computational Molecular Biology) and ISMB (Conference on Intelligent Systems for Molecular Biology) and this has caused a problem for researchers in computational biology. Some interdisciplinary conferences are sensitive to this issue (including ACM Electronic Commerce), and allow full papers to be submitted and reviewed but then published as a one-page abstract in the proceedings. But this is not the norm.

We offer the following vision of the future role of journals within CS, with some thoughts on how to make it come about. Many of these ideas have been suggested and, indeed, some have even been tried. But more serious experimentation is needed. Our vision of the future:

- Papers will be available on public web archives such as CoRR, the Computing Research Repository (see http://arxiv.org/corr), the CS part of the arXiv. This is increasingly common now, as researchers are discovering the advantages of posting papers on managed archives rather than just having them on their own home pages.

But if all papers are available in one place, then making a paper stand out from the pack will become more significant. One of the best ways of doing this will be via certification.

- Journals will be the main "certification" authorities, because they can operate without deadlines or strict page limits, allow for a careful review cycle with checking of results, are compatible with other scientific disciplines, and promote thoughtful work. (By "journal" here we simply mean an editor-in-chief together with an editorial board recognized by the community as a certifying authority." Journals like *Journal of Artificial Intelligence Research* (*JAIR*) and *Logical Methods in Computer Science* (*LMCS*) demonstrate that a group of community members can start viable, well-respected journals without the support of a publisher.) To paraphrase Churchill, we believe that journal reviews are the worst form of certification, except for all the other ones that have been proposed now that papers are available online. For example, citation counts (and page rank style variants of citation counts) suffer from well-known problems, including the fact that different fields have different citation rates, they can be influenced by fads, and the counts depend on the database of papers being used [4].[2] They cannot, for example, tell an economist that a computer science paper on game theory is relevant to economists, nor can they certify the correctness of results.[3] Nevertheless, citation counts are becoming increasingly important. An advantage of greater CS use of journals is that it would allow CS citations to be more comparable to those of other fields. This is particularly significant when CS researchers compete for, say, funding at the national level.
- For journals to play an important role in CS will require significant change. Journals must be much faster in reviewing papers, and indeed this will be essential in supporting the promising trend towards using journals also as a deadline-free path to conference presentation, in addition to protecting their more traditional role. Of course, de-emphasizing the importance of conferences will help to achieve this, by significantly reducing the work load on conference reviewers. Here are some additional suggestions for speeding up the review cycle:
    - It seems sensible to adopt a page limit for papers that require fast review, for example to facilitate presentation at an upcoming conference. The resulting process would still allow for a full review cycle and continuous submission throughout the year. (Having said that, we believe that it is critical that there be enough journals that do not have significant page limitations and allow for longer, thoughtful articles.)
    - For conferences that maintain their own review process, better coordination with journals would allow the same reviewer to read both the conference and an extended journal version.
    - More cooperation between journals would also be helpful. For example, a journal could agree to pass on reviews of paper and the names of the reviewers to another journal at the request of authors (subject to the agreement of the reviewers). Halpern followed this policy as editor-in-chief of the *Journal of the ACM*; authors of rejected papers could request reviews to be passed on to the journal of their choice.
- Each certification could come with a "cost": for every paper that is reviewed by a journal, some author of that paper must be available to review another paper (Crowcroft et al. [2]

make a similar point) and/or there could be a cost for submission. Both approaches are in fact used in the *B.E. Journal of Theoretical Economics* (see http://www.bepress.com/bejte/policies.html). Currently, authors can choose either to commit to reviewing two papers in a timely way (within 21 days of receiving it) or paying $350 when they submit a paper. If an author agrees to review a paper and his/her review is late, then there is a financial penalty (currently $200).

- Certifications need not be mutually exclusive. We believe that the community should experiment with different forms of certification. For example, those who work in interdisciplinary areas may choose to write two versions of a paper, targeted to different communities, and then get certification from the appropriate journals for each of the two versions. It may even make sense to get certifications from two (or more!) different communities for the same paper. [4]

The conference culture has served CS well up to now. Conferences provide authors with useful feedback; they're also a great forum for meeting colleagues. We should debate whether conferences should continue in the same role in the future, now that CS has matured and is making connections to so many other fields. As we have tried to make clear, the role of journals, and how certification will be carried out, needs to be an important part of this debate.

Acknowledgments: We thank Kavita Bala, Ken Birman, Lance Fortnow, Nir Friedman, Johannes Gehrke, Paul Ginsparg, Jim Hendler, John Hopcroft, Doug James, Christoph Koch, Jon Kleinberg, Daphne Koller, Steve Marschner, Michael Mitzenmacher, Andrew Myers, Greg Morrisett, Moshe Vardi, and two anonymous reviewers of the paper for their comments.

1. We remark that Fortnow's viewpoint is not universally accepted; see, for example, the discussion of different reactions in Vardi's recent Editor's letter [6].

2. A recent *Nature* editorial Assessing assessment" and a collection of metrics-related articles are available at http://www.nature.com/metrics.

3. Another certification approach that has been suggested is to have people just write reviews of papers, and attach them to the papers, without the need for recognized certification authorities." Again, while we believe that such reviews can play a useful role, we are not aware of any such system that has succeeded. Part of the problem is that the people whose reviews we would most like to read are busy; another is that a rather idiosyncratic set of papers will be reviewed this way.

4. We do not believe that copyright issues will present a serious impediment (any more than they are now in CS when different versions of a paper appear in both a journal and conference), and expect to see a continuing trend where authors do not give exclusive copyright to journals, giving them an assent to publication instead.